\documentclass{article}
\usepackage[margin=1in]{geometry}
\usepackage{url} 
\usepackage{tabularx,tabu}
\usepackage{hyperref}
\usepackage{booktabs} 
\usepackage{graphicx}
\usepackage[]{xcolor}

\AtBeginDocument{%
  \providecommand\BibTeX{{%
    \normalfont B\kern-0.5em{\scshape i\kern-0.25em b}\kern-0.8em\TeX}}}

\usepackage{graphicx}
\usepackage{hyperref}
\usepackage{ragged2e}
\usepackage{breakurl}
\usepackage{longtable}

\title{Computer Vision and Conflicting Values: \linebreak
Describing People with Automated Alt Text}

\author{Margot Hanley \ \ Solon Barocas \ \ Karen Levy
\\ \\
Shiri Azenkot \ \ Helen Nissenbaum \\}
\date{}

\begin{document}
\maketitle

\begin{abstract}
Scholars have recently drawn attention to a range of controversial issues posed by the use of computer vision for automatically generating descriptions of people in images. Despite these concerns, automated image description has become an important tool to ensure equitable access to information for blind and low vision people. In this paper, we investigate the ethical dilemmas faced by companies that have adopted the use of computer vision for producing alt text: textual descriptions of images for blind and low vision people. We use Facebook's automatic alt text tool as our primary case study. First, we analyze the policies that Facebook has adopted with respect to identity categories, such as race, gender, age, etc., and the company’s decisions about whether to present these terms in alt text. We then describe an alternative---and manual---approach practiced in the museum community, focusing on how museums determine what to include in alt text descriptions of cultural artifacts. We compare these policies, using notable points of contrast to develop an analytic framework that characterizes the particular apprehensions behind these policy choices. We conclude by considering two strategies that seem to sidestep some of these concerns, finding that there are no easy ways to avoid the normative dilemmas posed by the use of computer vision to automate alt text.
\\ \\
{\bf Keywords:} accessibility, alt text, visual impairments, computer vision, identity, gender, race, disability, Facebook, museums, policy
\end{abstract}

\section{Introduction}

Scholars have recently drawn attention to a range of controversial issues posed by the use of computer vision for automatically generating descriptions of people in images. These include the essentializing nature of classification, especially with respect to identity-related concepts like gender and race \cite{bowker2000sorting,scheuerman2019computers}; the presence of derogatory and offensive labels in training datasets \cite{hanley2020ethical,birhane2021large,crawford2019excavating}; and biases in labeling practice that negatively impact marginalized groups \cite{buolamwini2018gender}. Critics also warn of the privacy implications of such tools \cite{keyes2018misgendering,birhane2021large}.

Despite these urgent concerns, automated image description has become an important tool to ensure equitable access to information for blind and low vision (BLV) people. For many years, BLV people navigating the web have relied on \textit{alt text}---textual descriptions of visual content, attached to images as an “alt” attribute in HTML code. Alt text allows BLV people to have these descriptions read out to them by a screen-reader. Traditionally, alt text---when it is produced---has been produced manually and voluntarily by the person uploading the image. But in recent years, in the interest of addressing the paucity of user-provided alt text, platforms have turned to computer vision to automate this process, systematically assigning descriptions to images that otherwise might not receive them. A number of companies have embraced computer vision to improve the coverage and quality of alt text, including Google and Microsoft \cite{GoogleAccessibilityTeam, MicrosoftAutoDescriptions}. In 2016, Facebook announced its launch of “automatic alt text” (AAT), a feature that would provide its BLV users with descriptions of every photo on the platform \cite{wu2017automatic}. 

Creating descriptions of images, however, is not a straightforward task. The process of determining what information to include in a description is both technically difficult and ethically fraught. This is especially so when describing people; characteristics used to describe people’s identities change over time, and visual markers of identity are often tied to social constructs with troubling histories. Organizations creating alt text thus face difficult questions: Which features of an image are salient enough to merit description in alt text? How should they be described, and what can (or should) be omitted? What values should inform image description practices? Who benefits and who is harmed by different policies? And how sensitive should these determinations be to different contexts? 

When it released AAT, Facebook expressed normative concerns about many of these questions, and acknowledged the various trade-offs they entailed and open questions that remained \cite{wu2017automatic}. But Facebook is, of course, not the first organization to be confronted with versions of these questions. Museums have long navigated these tensions in their own practices of describing images in text, and have developed specific principles and guidelines to aid in their determinations, along with explicit justifications for their normative choices. To be sure, technology companies and museums are not completely analogous; the two face very different constraints in approaching the task of image description, and we don’t intend to suggest that platforms should (or could) adopt museums’ approaches wholecloth. Yet museums’ approaches are still instructive, revealing the fault lines that platforms must navigate in making their own design choices, and offering some strategies from which platforms might learn. In this respect, we situate this paper alongside scholarship like \cite{jo2020lessons}, which compares machine learning dataset collection to that of archives and calls “on the machine learning community to take lessons from other disciplines that have longer histories with similar concerns” \cite{jo2020lessons}.

We further aim to put our work into conversation with research by Facebook itself \cite{wu2017automatic} which acknowledges some of these tensions, as well as a line of work by scholars exploring the ethical stakes of image description and access by BLV users \cite{stangl2020person, macleod2017understanding, bennett2019point}. A notable contemporaneous work in this line of scholarship is \cite{bennett2021s}, which explores the preferences of BIPOC, nonbinary, and transgender users of screen readers, with particular focus on how race, gender, and disability are described in alt text. We aim to complement Bennett et al.’s rich qualitative work from users’ perspectives with our focus on comparative organizational practices.

We proceed as follows. We begin by reviewing the scholarly research on the normative dimensions of computer vision, image description, and accessibility for BLV users. We then present a study of the policies and normative justifications that two different types of organizations have invoked in deciding how to describe images---especially images of people---in text. First, we analyze the policies that Facebook adopts with respect to race, gender, age, and physical characteristics and the company’s decisions around whether to include or omit descriptions of this type from alt text. We then present an alternative---and manual---approach practiced in the museum community, focusing on how museums determine what to include in descriptions when the primary goal is to serve BLV people. We compare the similarities and differences between the policies adopted by Facebook and museums---and the expressed reasoning behind these policies. 

\section{Related Work}

\subsection{The production and prevalence of alt text}

Alt text online is infrequently available and of poor quality \cite{guinness2018caption, gleason2019s}. Responding to this problem, researchers have developed tools, strategies, and applications to increase its availability at scale. Some techniques to increase the quality and quantity of alt text are semi-automated; previous work has explored how to draw on crowd workers \cite{salisbury2017toward}, friends, and volunteers \cite{brady2015gauging} to provide image descriptions. These semi-automated systems have limitations. Highlighting the length of time it takes to provide descriptions to BLV people, for instance, researchers have noted that VizWiz, a crowdsourcing tool, takes more than two minutes to provide a description of a single image \cite{bigham2010vizwiz}. Researchers also report that relying on others can feel like a social burden \cite{rzeszotarski2014estimating} and describe the difficulty of implementing semi-automated systems at scale \cite{huang2016there}.

Recent work has focused on addressing these issues by creating more fully automated systems. These systems rely primarily on machine learning, computer vision, and natural language processing techniques. Researchers have advanced the general techniques of object recognition for automatic caption generation \cite {fang2015captions,karpathy2015deep,tran2016rich}. Recently, these have been applied specifically to the production of alt text, as in the case of Facebook’s AAT, which makes descriptions instantly available for all photos on its platform \cite{wu2017automatic}. Several companies have developed publicly available computer vision APIs that include object recognition and captioning, including Google Cloud, Microsoft Azure, and Amazon Rekognition \cite{GoogleCloudAPI, MicrosoftAzure, AmazonRekognition}.

\subsection{What users want from alt text}

Previous work has established that BLV people are interested in being presented with information about people in images \cite{zhao2017understanding} and that preferences about the type, quantity, and conditions under which information is presented varies across contexts. \cite{stangl2020person} explore the gap between what is typically offered in alt text and the stated preferences of BLV users. They find that BLV users want to be presented with detailed information about people in images but that preferences vary across contexts (e.g., dating websites, social media, news sources, etc.). On social networking sites, users want information such as the image subjects' gender, race, and physical appearance \cite{stangl2020person}. In other work, BLV users express that they feel it is more appropriate to describe people with concrete, visual details than with identity categories, although they qualify that in some contexts this information might be worth including, such as when users post photos where their race, gender, or disability is central to its meaning \cite{bennett2021s}.

Researchers have also found that BLV people want contextual information about people while navigating through public space; Branham et al. \cite{branham2017someone} find that BLV users would like to be able to identify people they know, the physical attributes of people, as well as demographic information. A small percentage of respondents said they wanted information that could allow them to meet strangers in emergency situations or find attractive strangers. This work affirms research indicating that preferences for information about people, including attributes like race and gender, vary based on the context in which the image is presented \cite{petrie2005describing,zhao2017understanding, bennett2021s, stangl2020person}.

\subsection{Normative implications of computer vision}

A rich line of scholarship has explored the social consequences of classification, an act which is inescapably political; categories map insufficiently onto the complexity of human experience and impose an external (and often rigid) viewpoint on who or what “counts” and who is sufficiently “like” someone else to be grouped together in a taxonomy \cite{bowker2000sorting, crawford2019excavating}. The acts of translation and categorization inherent to computer vision are likewise necessarily reductive and discretizing; they inevitably do some violence both to the richness and depth of visual media and the indeterminacy of identity \cite{hoffmann2020terms}. Classification of people at various stages of the pipeline leads to many types of normative concerns \cite{hanley2020ethical,kazimzade2020biased,yang2020towards}. The process of labeling images of people is inherently subjective \cite{miceli2020between, van2018talking, otterbacher2018social} and biased along dimensions of race and gender both of the image subject \cite{van2016stereotyping, barlas2019social, kyriakou2019fairness} and the labeler \cite{otterbacher2019we}. Unique challenges surface when the process of describing images---particularly images of people---is automated via computer vision. Prior work demonstrates that computer vision disproportionately misidentifies or fails to identify members of marginalized communities, such as people of color \cite{buolamwini2018gender} or non-binary people \cite{scheuerman2019computers,keyes2018misgendering}, and reduces gender to a stable, binary classification system \cite{hamidi2018gender, scheuerman2019computers}. Other scholars have remarked on the dangers of using computer vision to draw inferences about personal characteristics that are not visually evident \cite{yang2020towards, van2016stereotyping}, including sexual orientation and criminality \cite{wang2018deep,hashemi2020criminal}. The output of computer vision models can reinforce harmful stereotypes \cite{barocas2017problem}, subject people to offensive description \cite{crawford2019excavating,birhane2021large} and otherwise raise serious questions about representational justice \cite{hoffmann2018data}. 

These areas of research demonstrate how organizations implementing automated alt text must balance multiple interests. On one hand, organizations may seek to improve accessibility by providing the rich information desired by BLV users; on the other, they may want to do so in a way that attends to concerns about the dangers of using computer vision for describing people. In what follows, we consider how these tensions are instantiated in the alt text practices and policies developed by two very different types of organizations.

\section{Methods}

To illustrate how these tensions are confronted in practice, we analyze policies related to describing people in alt text in two types of organizations: Facebook’s AAT system and museums’ guidelines for composing alternative descriptions of images in their collections. We focus on the Museum of Contemporary Art in Chicago (MCA) and Cooper Hewitt Smithsonian Design Museum in New York City, each of which has played a prominent role in developing alt text guidelines in the museum community. We describe the policies each organization has adopted for producing alt text, then interrogate and compare the normative reasoning underlying these design choices, to the extent we can derive it from organizations’ explanations of their goals and rationales.

Facebook does not publish its policies about how it describes images of people in AAT; museums make their guidelines publicly available. Thus, our analysis involved first identifying all available information related to Facebook’s policies for describing images of people with AAT and, where available, its stated rationales for these policies. We conducted a systematic review of public sources describing Facebook's AAT tool, including Facebook’s published research paper on the topic \cite{wu2017automatic}, the company website and blog posts, and broader media coverage. Our search involved querying Google News and Factiva for articles including the terms “Facebook” and either “automated alt text” or “automated alternative text,” as well as site-specific Google searches of Facebook’s website and blog posts. We read each result and noted any mention of Facebook’s policies or rationale around describing images of people with AAT. Altogether we reviewed 539 responsive results; though the vast majority of these were duplicative, this review gives us reasonable confidence that we have comprehensively surveyed the landscape and located Facebook’s public statements about AAT.

Importantly, our analysis here focuses on publicly available sources. We did not formally audit Facebook’s AAT system, nor secure “insider” knowledge about how its alt text practices were determined, how they are put into practice, or the rationales justifying those choices. In both cases, our analysis centers on stated policies rather than enacted practices, since our primary interest is in organizations' expressed normative reasoning (rather than specific workings of the systems in practice). 

\section{Automatic Alt Text at Facebook}

Facebook introduced AAT in 2016, with the goal of ensuring that BLV users could “get the same enjoyment and benefit out of the platform as people who can [see photos]” \cite{wu2017automatic,VergeFacebook}. In their previous research, the company found that BLV users felt frustrated or isolated by their experience engaging with images on the platform \cite{zhao2017effect,FbookEngBlog}. While algorithmically generating alt text for every user-uploaded image on the platform posed an extremely difficult technical challenge, the accessibility team pursued the effort as one core to the company’s mission, asserting that accessibility is a requirement of “connecting the world” \cite{Wapo} and emphasizing the team’s ethos that what “visually appears for sighted people gets translated well into something that’s non-visual” \cite{TechCrunchFacebook}. 

While media coverage of AAT was positive, highlighting the Facebook’s accessibility team's promotion of inclusivity and equity \cite{VergeFacebook,Engadget,Wapo}, feedback from the BLV community was mixed, largely due to the sparsity of detail in image descriptions. Accessibility scholar and advocate Chancey Fleet has noted that Facebook’s AAT is “famously useless in the blind community” despite “garner[ing] a ton of glowing reviews from mainstream outlets” \cite{Chancey}. 

There was also criticism from the broader public, reflecting expectations that Facebook should provide richer descriptions. A platform-wide outage on July 3, 2019, exposed AAT’s shortcomings to sighted users who had not previously encountered the feature \cite{VergeOutage}. During the outage, Facebook users could access the platform but could not upload or see images. Instead, they saw blank white boxes with small, blue alt text describing the images that were not displayed. This led to a flurry of posts on Twitter in which Facebook users commented on the poor quality of the AAT, noting its vagueness and questioning whether it constituted an “inclusive experience” \cite{annatweet,MykalTweet}. In his 2017 exhibit “Image May Contain,” the artist-researcher Lior Zamalson highlighted the tool’s shortcomings by illustrating how AAT describes famous photographs of historical significance, showing how it “flattens [images’] meaning by taking them out of their social context” \cite{LiorTheFix}. For example, the image of President John F. Kennedy “driving through the streets of Dallas in a Lincoln Continental just moments before he was assassinated” was translated by AAT to “Ten people, car.” Reflecting on the exhibit, Fleet noted how the descriptions are “stripped of valence” \cite{TWIMLAI}.

The criticisms can be attributed, in part, to technical limitations, recognized by Facebook itself, such as choosing to restrict alt text to tags rather than generating rich and fluid captions, as the system is “limited by the accuracy and consistency of existing natural language caption generation systems” \cite{wu2017automatic}. Furthermore, Facebook notes the challenge of applying existing algorithms for caption generation to AAT’s design and implementation; these models, which were "designed and evaluated for sighted people" \cite{wu2017automatic} tend to identify and describe objects more relevant for sighted users than BLV users. Ultimately, some of the public’s criticism---which highlighted how flat, brittle, and unnatural the descriptions read---can be explained by these technical challenges. 

Other aspects of Facebook’s AAT design, however, stem from the company’s policy decisions  dictating what descriptions should contain. These policy decisions are distinct from the system’s technical constraints: even if a company could theoretically generate long, vivid automatic descriptions, it would nonetheless face the question of what concepts should be described, and how. These policy decisions are inescapable: other companies with comparable object detection services apply policies of their own around which identity attributes they will return in image description and alt text. In 2020, Google adjusted its Cloud Vision API so that it would no longer return tags for gender, such as `man’ or ’woman,’ but would instead return “non-gendered label[s] such as ‘person’ […] given that a person's gender cannot be inferred by appearance” \cite{GoogleGender}. Microsoft, which offers image captioning services to “improve accessibility” through Azure AI, appears to include gender in some instances and not others \cite{MicrosoftAzure}. As of April 2021, Amazon Rekognition makes “gender binary (male/female) predictions based on the physical appearance of a face in a particular image” \cite{AmazonRekognition}.

\subsection{Facebook’s AAT policies}
The primary means through which Facebook operationalizes its AAT policies is through the use of a blocklist to prevent alt text from including information about certain “sensitive” categories. Facebook applies this blocklist to a set of pre-defined prominent and frequent concepts. Lastly, Facebook employs an “accuracy threshold” below which it will not return information in certain categories. These processes, detailed in \cite{wu2017automatic}, are elaborated below.

Acknowledging that there is an “infinite number of details in a given photo,” Facebook first had 30 human annotators review a subset of 200,000 randomly selected photos and then select three to ten “prominent” concepts per image. It then sorted the concepts by frequency and chose the top 200 as “concept candidates.” It then filtered those concept candidates to exclude those that “could have a disputed definition or that were hard to define visually,” like “concepts associated with personal identity,” including gender-related concepts (to which we return below); “fuzzy and context-dependent” adjectives like `young’ and `happy’; and concepts that are difficult for an algorithm to learn, like `landmark’ \cite{wu2017automatic}.

Applying this blocklist to the list of 200 concept candidates resulted in the removal of 103 concept candidates, leaving 97 concepts total. Importantly, this list purposefully excludes concepts not because they were not prominent or frequent, but due to other considerations. For example, the company considers gender an identity attribute and will categorically omit it from AAT descriptions. While the company does not outline explicit policies for describing age or physical appearance (other than that `young’ is an excluded “fuzzy or context-dependent” adjective), final concepts do include age-related words like `baby’ and `child’ and concepts which could be considered gender proxies, like `beard’ and `jewelry’ \cite{FbookEngBlog}.

In addition to categorical omission of some sensitive attributes via the blocklist, Facebook adopts a policy regarding the confidence threshold for predicting the concept accurately, which further distinguishes concepts by sensitivity. The company states that the algorithm must be at least approximately 80\% confident about its classification before it would display a concept in AAT \cite{wu2017automatic}. When it isn’t confident, “Facebook simply won’t suggest a description”; as the project lead notes, “[i]n some cases, no data is better than bad data” \cite{VergeFacebook}. In its research paper, the company does not mention its policies on ethnicity or race. However, in other sources, a Facebook representative has stated that AAT might return race if the model had sufficient confidence: “in sensitive cases---including ones involving race, the company […] will require a much higher level of confidence before offering a suggestion” \cite{VergeFacebook}. Facebook’s research paper notes that the requisite confident level is set at high as 98\% for “a few more sensitive concepts” \cite{wu2017automatic}.

In January 2021, Facebook announced that it was increasing the number of concepts that it includes in AAT from 200 to 1200, but the company has not commented on any changes in its policies around blocklisted or sensitive concepts, like gender or race. Facebook also announced that its new model is trained to perform consistently across the dimensions of skin tone, age, and gender \cite{facebookai2021}. In its 2017 research paper, Facebook stated that at that time it chose not to implement facial recognition until it could responsibly assess the privacy implications of doing so, despite requests for such a feature from research participants \cite{wu2017automatic}. Since then, the company has introduced the tool across the platform as an opt-in feature, and incorporated it into the AAT functionality \cite{facebookai2021}. 

\subsection{Normative justifications}
In its research paper and public statements, Facebook explicitly acknowledges a set of normative concerns and considerations that reflect the interests of various parties---including the subjects depicted in the image, the BLV user receiving the image description, and the person who uploaded the image \cite{facebookai2021, wu2017automatic, VergeFacebook}. 

Regarding image subjects, Facebook expresses concern around offensive mistagging or essentialization. For example, Facebook’s paper notes that while participants in users studies desired more detailed descriptions of image subjects, Facebook was hesitant to do so because of “the consequence[s] [of] wrong or potentially offensive tags” \cite{wu2017automatic}. By categorically omitting gender because “gender identification is more complex, personal, and culture-dependent than what may appear visually on photos,” the authors express concern about mistagging or essentializing image subjects. The authors acknowledge that in addition to inherently offensive concepts, there are also concepts which may become offensive when assigned to an object; they use the example of tagging a `person’ as a `horse.’ The privacy interests of image subjects are also implicated in Facebook’s expression of hesitation regarding the integration of face recognition into the AAT tool. The authors frame this concern as a tradeoff against BLV users' desire for more informative alt text.

As to BLV users, the authors highlight concerns around providing incorrect or possibly misleading information about an image, impeding effective social interaction or even resulting in BLV users being led to believe something embarrassing or offensive about the people depicted \cite{wu2017automatic}. The authors frame the latter problem as a "value/risk" tradeoff, expressing the concern that the cost of “algorithmic failure” (i.e., inaccurate tags) is uniquely high in the case of alt text. BLV people cannot visually assess whether a description is inaccurate, and the misunderstanding could lead a BLV person to make “inappropriate comments'' about other users’ photos \cite{wu2017automatic}. The authors specifically refer to the now-infamous situation in which Google’s image tagging feature misidentified Black people as `gorillas’ \cite{Googleapologises}, and express concern about the prospect of misleading BLV users with such an error; they cite this concern as justification for the high levels of confidence required before displaying certain image tags. 


Finally, Facebook expresses some concerns around the interests of the person who uploaded the image to which AAT is being applied (whom it refers to as the “photo owner”). In discussing the possibility of giving image uploaders the capacity to review and verify possible descriptions before making them available via AAT, Facebook expresses concern that doing so may “create a feeling of social debt for blind users” and that such a process could result in “a significant amount of work” for image owners, given that many photos will not be consumed by AAT users \cite{wu2017automatic}. Facebook also expresses concern about whether applying alt text to an image uploader’s photo could undermine their agency or “creative ownership” over an image. 

\section{Manual Practices in Museums}
                            
In light of the challenges Facebook encounters in its policy choices around AAT, we consider a separate context with a well-developed tradition of attaching text descriptions to images: museums. Museum practices impart a unique perspective in how they implement appropriate policies for alt text, and provide an important point of comparison for Facebook’s practices. We reviewed the image description guidelines for two museums, MCA and Cooper Hewitt, each of which has garnered attention for its alt text practices.

As an initial matter, it is notable that museums may offer multiple versions of textual descriptions for a single image; Cooper Hewitt offers both short and long descriptions for orienting BLV people to exhibits and artworks. While screen readers default to providing the short description, guests can opt into being read a longer and more detailed description \cite{Artsy}. Here, we consider both types.

An explicitly articulated set of normative values underpins museums’ image description practices. Cooper Hewitt presents a guiding philosophy for how it approaches composing descriptions of images which include people: 

\begin{quote}
``One should identify the visual appearance of persons, especially when it is important to the understanding of the content of an image. Instincts for political correctness tend to inadvertently result in redacting information; it is a general rule that if you are able to see something it should be shared with those who cannot see it'' \cite{CHGuidelines}.
\end{quote}

\noindent Cooper Hewitt’s guidelines recommend that all visual information should be available to BLV people in text descriptions. This principle is one often shared by individuals within the accessibility community. Sina Bahram, an accessibility consultant who advised both MCA and Cooper Hewitt on their guidelines, emphasizes the importance of rendering visual information explicit in certain contexts: 
\begin{quote}
``When I think of a picture or a photo of somebody using a wheelchair, I'm not necessarily interested in identifying disability. It would be, however, critical, in certain contexts to know that that person is using a wheelchair and in other contexts it may not be of interest, but I don't want to err on the side of, `Oh, we shouldn't talk about this because it's difficult.' Because I think that leads us to implicit censorship, and that's not what we want to do'' \cite{TWIMLAI}.
\end{quote}

Museums codify this philosophy in guidelines that offer recommendations for describing people in alt text. Both museums suggest describing an individual as a `person’ without attributed gender, unless gender is “clearly evident and verifiable” (MCA) \cite{MCAGuidelines} or “clearly performed and/or verifiable” (Cooper Hewitt) \cite{CHGuidelines}. The “or” in Cooper Hewitt’s guidelines would seem to allow for the use of gender even if not verifiable if it is “clearly performed”—and this is well reflected in the examples included in the guidelines, which use gender throughout. Similarly, both museums recommend avoiding any racial or ethnic terms unless race or ethnicity is “obvious and known”; in keeping with the guidance to rely on visual information, they instead recommend describing skin tone. Cooper Hewitt’s guidelines note that “where skin tone is obvious, one can use more specific terms such as black and white, or where known and verified, ethnic identity can be included with the visual information: Asian, African, Latinx/o/a (also see gender), etc."~\cite{CHGuidelines}. Bahram elaborates: ``I think that it is important to identify what is visually apparent in images when we're working with institutions on describing photos in the art context. We describe skin color, but we don't describe ethnicity because one can be seen, and the other one is inferred"~\cite{TWIMLAI}. He emphasizes this critical distinction between what information “can be seen” and which cannot; skin color is visually perceived and therefore should be described, whereas a person’s ethnicity, unless clearly represented through some other contextual cue, is not purely visual information and is therefore not amenable to description.

While museums recommend explicit guidelines for composing image description, they also acknowledge that these guidelines are a “living document” and must be constantly revisited to “engage with contemporary dialogues” \cite{CHGuidelines}. They emphasize the importance of “regularly review[ing] and updat[ing] these guidelines and glossaries of terms to sensitively describe people without implications of judgement” noting that “image description inherently intersects with questions of race, gender, and identity” \cite{CHGuidelines}.

\section{Comparing Policies}

There are a number of similarities and differences to note about the policies adopted by Facebook and museums---and some of the expressed reasoning behind these policies. We summarize these in Table \ref{tab:commands} and highlight key points below.

Facebook and museums both appear willing to use race tags under certain circumstances, but rely on different justifications for doing so. Media coverage suggests that Facebook would tag race if its model is sufficiently confident, with the confidence threshold set especially high \cite{VergeFacebook}. In contrast, museums seem to endorse its use only when it is known or verified. Museums are more permissive when it comes to skin tone, however: when it is clearly visible, they suggest that it should be noted in alt text. Nearly all of the alt text for example images in Cooper Hewitt’s guidelines includes details about skin tone. Facebook, to our knowledge, has made no public statements about its willingness to describe skin tone.

Facebook and museums depart substantially when it comes to their policies on gender description. Facebook categorically refuses to use gender terms. Museums, in contrast, endorse their use under certain conditions: Cooper Hewitt notes that gender can be named if “clearly performed and/or verifiable,” while MCA only allows its use if “evident and verifiable.”

Museums seem to embrace the use of age descriptions with no obvious restrictions, while Facebook purposefully removed `young’ from AAT’s set of available concepts, suggesting that it hesitates to infer people’s age. Facebook does use certain concepts that still indicate age (e.g., `child’ and `baby’); notably, unlike `young,’ these do not seem to be filtered out as part of its process of removing tags with ``disputed definition or that [are] hard to define visually'' \cite{wu2017automatic}.

Cooper Hewitt is unique in suggesting that alt text should describe apparent physical disabilities. Note, however, that in its examples, Cooper Hewitt’s alt text does not use the term `disabled’ or name a specific physical disability; the examples instead mention the appearance of a wheelchair. It is thus unclear whether Cooper Hewitt endorses using the language of disability or prefers that alt text only describe observable properties like assistive devices. The MCA guidelines do not remark on disability, nor does Facebook.

Both Facebook and museums, however, seem committed to describing physical features of particular importance, including those that might serve as the basis for inferences about identity categories like gender and race. For example, Facebook gives `beard’ and `jewelry’ as examples of concepts included in AAT; the company has also suggested that it would like to include attributes like hair color \cite{Engadget, TechCrunchFacebook, MashableFacebook}. As mentioned, both museum guidelines advocate in favor of describing skin tone, which can clearly serve as the basis for inference about race; MCA's guidelines even mention hair color. 

Finally, while both museums advocate in favor of identifying recognizable (i.e., public, historical, etc.) figures by name, Cooper Hewitt says that doing so does not obviate the need to also describe the person’s physical features. Its example of alt text appropriate for a painting of Michelle Obama names her, but also mentions her skin tone. Facebook originally hesitated to use facial recognition to identify people by name---either when describing users’ friends or even public figures. Matt King, an Accessibility Specialist at Facebook, even expressed frustration that "[i]f everyone else can see that's a picture of Donald Trump, I should be able to know that too" \cite{Wapo}. The company has since changed its policy to allow for the use of facial recognition if users that appear in photos have opted into the feature. It is unclear, however, whether Facebook reports any other details about people if they have been named.

\section{Axes of difference}
\subsection{Manual versus automatic alt text}

Can the differences in Facebook's and museums’ policies simply be explained by the fact that one relies on manual annotations by humans and the other on automatic descriptions based on computer vision---under the assumption that the normative implications of each technique are so distinct as to demand different policies?

Humans can be more discerning than computer vision models---considering a wider range of details in an image, taking context into greater account, or relying on other expert knowledge. Humans have the capacity to reflect critically on their assumptions, to recognize their own uncertainty, and to decide when to seek out more information to make an accurate determination. They may also have the means to do so, perhaps through contact with the person whose identity is in question or by locating sources in which the person has shared details about their identity. Humans will be aware of the normative implications of making certain kinds of mistakes (e.g., describing a person as an animal). And humans can even decide to abstain from offering descriptions given their uncertainty, or decide to communicate that uncertainty in the descriptions themselves---something that might be especially important when further information is unavailable.


In practice, however, both manual and automatic approaches might still result in people being described in terms not of their own choosing. Museums could just as easily subject a person to gender or racial classification as Facebook’s AAT---and indeed examples of alt text for many images in the museum guidelines suggest that they are willing to do so in some circumstances. That museums might be more accurate or more circumspect in describing the gender or race of a person than Facebook does not sidestep the normative objections to engaging in gender and racial classification in the first place \cite{keyes2018misgendering,hamidi2018gender}. Notably, many of the findings from Bennett et al.’s interviews that express concerns about ascribing characteristics to people were not limited to automated alt text; they applied equally well to manual alt text \cite{bennett2021s}. Furthermore, humility---in the form of recognizing and communicating uncertainty---is not the preserve of human annotators. Computer vision models can be designed to quantify the certainty of their inferences. As mentioned earlier, Facebook spent a good deal of time using such quantifications of uncertainty to decide what should serve as a minimum threshold of confidence to report certain inferences. In fact, Facebook even experimented with reporting uncertainty along with each tag in the alt text, but their study revealed that users found such information “cumbersome and hard to interpret” and thus the company decided to omit it \cite{wu2017automatic}. The company instead starts each alt text with “Image may contain” to stress uncertainty, generally \cite{FbookEngBlog,wu2017automatic}. Thus, the differences in Facebook and museums’ policies are not entirely explained by the fact that the former relies on automatic descriptions, while the latter relies on manual descriptions. 

\subsection{Social media versus museums}

Social networking sites and museums serve very different functions in society. Perhaps it is unsurprising that the normative issues at stake when producing alt text are also very different across these organizations---and that their policy positions differ accordingly. Online social networks are places where people go to connect with others, but also cultivate their identity and express themselves \cite{bennett2021s}. Museums are cultural institutions that aim to curate important artistic contributions or cultural artifacts, make them available to broad audiences, and provide some interpretation and context to help patrons appreciate their meaning and significance. Facebook may naturally want to stake out a policy position that reinforces the view that it is merely a neutral conduit for social interactions among peers. In contrast, the very nature of museums is that they choose where to direct patrons’ attention and that they play an active role in shaping the interpretation of displayed works. Even though the Cooper Hewitt guidelines caution that “[b]ringing interpretive knowledge to a description is not always preferred” when crafting alt text---encourage annotators to focus on what is visually evident and deferring interpretation to patrons---the museum still allows that there may be circumstances “when interpretive knowledge aids visual understanding of the content [and that] it should be incorporated” \cite{CHGuidelines}. No such imperative seems to exist at Facebook.

\begin{table*}
{\footnotesize
\centering
  \caption{Policies for Describing People by Organization}
  \label{tab:commands}
\resizebox{0.99\textwidth}{!}{%
\begin{tabular}{p{0.08\linewidth}p{0.22\linewidth}p{0.33\linewidth}p{0.30\linewidth}}
    \\
    \toprule
    \multicolumn{1}{c}{\hspace{1cm}} & \multicolumn{1}{c}{\textbf{Facebook}} &
    \multicolumn{1}{c}{\textbf{Cooper Hewitt}} &
    \multicolumn{1}{c}{\textbf{Museum Contemporary Art Chicago}}
     \\
    \midrule
\\ \textbf{Race/\hspace{1cm}
Ethnicity/\hspace{1cm} Skin Tone} & 
``By default, Facebook will only suggest a tag for a photo if it is 80 percent confident that it knows what it’s looking at. But in sensitive cases---including ones involving race […] it will require a much higher level of confidence before offering a suggestion. When it isn’t confident, Facebook simply won’t suggest a description."~\cite{VergeFacebook}   & 
``When describing the skin tone of a person use non-ethnic terms such as ``light-skinned'' or ``dark-skinned'' when clearly visible. Because of its widespread use, we recommend the emoji terms for skin tone as follows: \includegraphics[height=1.1em]{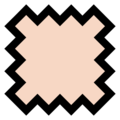} Light Skin Tone, \includegraphics[height=1.1em]{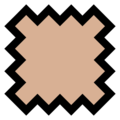} Medium-Light Skin Tone, \includegraphics[height=1.1em]{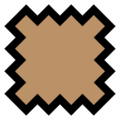} Medium Skin Tone, \includegraphics[height=1.1em]{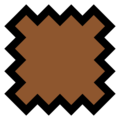} Medium-Dark Skin Tone, \includegraphics[height=1.1em]{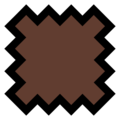} Dark Skin Tone. Also, where skin tone is obvious, one can use more specific terms such as black and white, or where known and verified, ethnic identity can be included with the visual information: Asian, African, Latinx/o/a (also see gender), etc.''~\cite{CHGuidelines}   & 
``Demographic: race. This is in development, but for the time being identify clearly visible visual appearance when it is important to the understanding of the content. Default to ``light-skinned'' and ``dark-skinned,'' when clearly visible. Where obvious and known, use more definite terms; e.g. black, Latino, Asian, etc.'' \cite{MCAGuidelines} \\ \\ 
\hline
\\ \textbf{Gender} & 
``[W]e decided to leave out gender-related concepts such as woman/man, girl/boy, as gender identification is more complex, personal, and culture-depdendent than what may appear visually on photos.'' \cite{wu2017automatic} & 
``No assumptions should be made about the gender of a person represented. Although, where gender is clearly performed and/or verifiable, it should be described. When unknown, a person should be described using ``they, them'' and ``person'' and their physicality expressed through the description of their features, which inadvertently tend to indicate masculine or feminine characteristics. The use of masculine and feminine are problematic and should be avoided unless necessary for describing the performance of gender.''  \cite{CHGuidelines}  & 
``Where necessary for understanding content gender may be described, but no assumptions should be made. Our default should be ``person'' except where gender is clearly evident and verifiable.'' \cite{MCAGuidelines} \\ \\
\hline
\\ \textbf{Age} & 
``We ended up with a list of 97 concepts […] including people (e.g., people count, smiling, child, baby)'' \cite{wu2017automatic} & 
``Describe the age of represented people in an image using terminology such as baby, toddler, child, youth, teen, young person, adult, older person.'' \cite{CHGuidelines} & 
``Use terms that indicate age: baby, toddler, child, youth, teen, young adult, adult, older person.'' \cite{MCAGuidelines} \\ \\
\hline
\\ \textbf{Disability} & 
 & ``Not only […] prominent features or physical stature, but also physical disabilities [should be described].'' \cite{CHGuidelines} & 
  \\ \\
\hline
\\ \textbf{Physical Features} & 
``The current list of concepts covers a wide range of things that can appear in photos, such as people’s appearance (e.g., baby, eyeglasses, beard, smiling, jewelry)'' \cite{FbookEngBlog} & 
``When particular features are immediately noticeable, or mutually agreed upon salient features of a known person are visually present, they should be described.'' \cite{CHGuidelines} & ``need to create reference list for […] hair color'' \cite{MCAGuidelines} \\ \\
\hline
\\ \textbf{Identity} & 
``And since people who use Facebook often share photos of friends and family, our AAT descriptions used facial recognition models that identified people (as long as those people gave explicit opt-in consent).''~\cite{facebookai2021} & 
``When describing an image of a recognizable person, identify them by name, but also describe their physical attributes. If an individual is not a public figure, and the context does not imply the importance of who is represented, it may not be appropriate to identify the individual.'' \cite{CHGuidelines} & 
``Feel free to identify clearly recognizable figures, e.g. Jesus, Bozo the Clown, Madonna, Anne Kaplan, and Sammy Davis Jr.'' \cite{MCAGuidelines} \\ \\
    \bottomrule
  \end{tabular}}}
\end{table*}

The structural positions of stakeholders also differ across these organizations. On Facebook, users are often both the producers and consumers of visual media---and thus hyper-attuned to presentations of the self and others; in museums, patrons are almost always only consumers of the displayed works. The purpose of alt text in these contexts is thus very different: in one, to facilitate social interaction among people in structurally similar positions; in the other, to facilitate lay appreciation of an expertly curated set of cultural artifacts. If patrons’ \textit{own} identities are at stake in the alt text generated in museums, it is only by virtue of how museums describe \textit{others} that belong to the groups with whom patrons may identify. Though museums are surely attuned to the concerns and interests of artists and the creators of other artifacts---as well as the possibly real people represented in these works---such concerns are different than those museums have for their patrons.

There are also many \textit{practical} reasons why Facebook and museums have adopted quite different practices and why these might account for notable policy differences. Facebook may feel compelled to adopt automated methods, given the scale of the task, but also be more conservative in its use of identity terms, given that computer vision models will likely perform less well than humans and fail to engage in the types of deliberation described above. Facebook has committed itself to generating alt text for all images on its platforms, the scale of which is far larger than any collection that a museum might attempt to annotate. Three hundred and fifty million images are uploaded to Facebook each day \cite{FacebookStats}; 
the MCA’s collection stands at 2,500 pieces \cite{Collection}. Facebook generates alt text for visual content as soon as it has been uploaded to the site; museums have the comparative luxury of taking some time to catalog a new piece in their collection, including adding alt text before making the piece publicly available. For Facebook, manually producing alt text at this scale and speed is likely infeasible---and certainly would be if the goal were to produce alt text of the type produced by museums.

\section{Sources of uncertainty}

The goal of comparing Facebook and museums’ policies is not to suggest that they should be identical, or that one is clearly preferable to the other. There are many good reasons why they differ. Yet not all of these differences can be explained away by the fact that they adopt different techniques (one manual, the other automatic) to generate alt text or that they serve very different social functions. In drawing out these differences, we have tried to throw into greater relief the range of choices and normative considerations that go into alt text, whether explicitly or perhaps unwittingly. In what follows, we offer a deeper analysis of the normative reasoning that might---and should---undergird these choices.

As we have described, Facebook’s statements evince a range of justifications for its decision-making around tagging identity categories in AAT. Several related, but analytically separable, sources of potential uncertainty emerge, which we might group into three categories: (a) uncertainties related to technical accuracy; (b) uncertainties related to the ontology and epistemology of social categories; and (c) uncertainties related to social context and salience.

\textbf{Limits on technical accuracy.} The first source of uncertainty relates to limitations on computer vision's ability to reliably classify images. A good deal of Facebook’s discussion of identity categories invokes technical thresholds that must be met before it may display certain identity-related terms. As described, heightened confidence levels---up to 98\%---are required for particularly ``sensitive'' concepts, seemingly including race \cite{VergeFacebook, wu2017automatic}. Through these restrictions, Facebook imposes a requisite level of confidence that its algorithm is \textit{accurately} classifying images according to technical criteria.

The heightened accuracy required for certain identity-related concepts suggests that Facebook recognizes that erroneous tags around these concepts can cause particular harm: it may be more troubling to have a sensitive personal characteristic mistagged than, for example, to have Facebook mistake an apple for a pear.
This concern is further reflected in Facebook’s discussion of the relative risks of returning no tags for an image versus returning inaccurate tags for an image (balancing precision and recall): as Facebook notes, “in some cases, no data is better than bad data” \cite{VergeFacebook}. Ensuring a higher degree of technical accuracy, then, is viewed as a partial means of mitigating the potential harms of AAT.

\textbf{Ontological and epistemological limits of social categories.} A quite different source of uncertainty relates to the nature of the characteristics that comprise social identity. Rather than being stable, discrete categories with fixed meanings---as they are often treated in computer vision models---race, gender, disability, and other identity categories are complex social constructs, the meanings of which are fluid and contextual. As many scholars have noted, identity categories are unstable, historically and politically contingent, and premised on social hierarchies; they cannot be merely affixed to individuals as stable biophysical attributes \cite{benthall2019racial, scheuerman2020we, whittaker2019disability}.

A related concern is epistemological. Even if we \textit{were} to treat race, gender, age, disability, or other identity categories as fixed categories, we would still be limited in what can be learned \textit{visually} about these attributes. While certain markers of gender and race may be expressed visually (as Benthall and Haynes describe it, the category of race is partially based on “the assigning of social meanings to … visible features of the body” \cite{benthall2019racial}), they encompass a much broader range of nonvisible dimensions, and cannot be reduced to externally visible phenotypic characteristics. Many disabilities are “invisible”---that is, not readily apparent from physical appearance---which can itself impose particular burdens of recognition, legitimation, and access \cite{davis2005invisible}. As Kate Crawford and Trevor Paglen write, computer vision tends to rely on the assumption that concepts are not only themselves “fixed, universal, and have some sort of […] grounding and internal consistency” but also that there are “fixed and universal correspondences between images and concepts” such that “the underlying essence [of a concept] expresses itself visually” \cite{crawford2019excavating}. Both of these assumptions often fail---and fail consequentially---in the context of identity categories.

Some of Facebook’s policies allude to these ontological and epistemological constraints---for instance, its decision to omit gender tags from image descriptions because “gender identification is more complex, personal, and culture-dependent than what may appear visually on photos” \cite{wu2017automatic}. This reasoning seems to evince a hybrid recognition of both ontological and epistemological constraints, noting that gender is neither a construct that can be consistently applied from person to person, nor one that can be readily identified visually. A similar logic seems to undergird Facebook’s decision to redact ``fuzzy and context-dependent'' adjectives like `young' \cite{wu2017automatic}: what does `young' mean? What are the boundaries between `young' and `not young' and how readily can `young-ness' be inferred from a photo? Facebook’s policies seem to acknowledge that there may be more consistent agreement at one pole of a malleable identity concept---about the `young-ness' of, say, a `baby,' a tag which Facebook \textit{will} return---but avoids tags where line-drawing is made more difficult due to the nature of the construct and its (in)visibility.

\textbf{Social context and salience.} Finally, normative considerations around alt text relate to uncertainties related to social contexts, and what features are useful or appropriate to mark or make explicit within them. At least two contexts are at issue. The first involves what is being depicted in the photograph being described by AAT. Race, gender, age, disability and other identity categories may be central to understanding the social meaning of an image. The museum guidelines make this explicit, clarifying that visual appearance should be described especially to the extent that it is important to understanding an image’s meaning. Zalmanson’s work underscores the point in his description of the photo of the Kennedy assassination as ‘ten people, car‘: the absurdity of the caption stems from its remove from the social meaning and historical import of the image. For some images, the race and gender of depicted people are clearly salient, and key to understanding what the image conveys; for others, it is incidental, or might even be offensive or essentializing to make reference to them. But unlike manual practices in museums, which are informed by knowledgeable human annotators who can consider the meaning of each piece of art individually, Facebook’s AAT tool is not equipped to make such nuanced judgments. Facebook did consider the ``salience'' of different features of an image in constructing its initial list of concepts, in that it asked human annotators to tag a random selection of 200,000 photos publicly posted to Facebook with a limited number of tags (i.e., the most salient tags for the image), the top 200 most common of which were added to the initial concept candidate list \cite{wu2017automatic}. But this is a rather limited notion of salience in the context-specific sense in which we intend it here: while used to construct a list of concepts to be applied globally, it cannot address the specificities of how these tags contribute to the social meaning of particular images.

An independent consideration related to salience is the relational context in which the BLV user is consuming the image. The relationships among the BLV user, the depicted image subject, and the photo uploader are complex and variable. It may or may not be useful or appropriate for Facebook to call out identity concepts within different relational contexts. Bennett et al.’s \cite{bennett2021s} interview subjects describe how knowing the identity categories of the people with whom they interact on social media can in some cases be useful to facilitate code-switching and creating a base level of understanding for community-building (though acknowledge that doing so often involves reliance on assumptions). In other relational contexts, identity categories are less salient, and it may be irrelevant or inappropriate to make explicit reference to them. Facebook’s introduction of face recognition into its AAT model (as well as the identification of image subjects by name) seems geared toward addressing the relevance of relational context. Facebook also relies on this reasoning in explaining its recent decision to permit users to expand an AAT image description into additional detail, noting that screen reader users “[want] more information when an image is from friends or family, and less when it’s not” \cite{facebookai2021}.

To illustrate how these sources of uncertainty apply to image descriptions, consider the question of whether an AAT system should return the tag `wheelchair.' We might reasonably assume that it is feasible for a model to accurately tag most images of wheelchairs with high confidence, surpassing the first hurdle (technical accuracy). We might also assume that we can reasonably delimit a category of objects that comprises `wheelchairs' and that we can detect visually what objects belong to such a category (addressing ontological and epistemological concerns). `Wheelchair' is in these senses an easier case than `woman' or `young' or `Asian'---and it might seem reasonable, then, to include a `wheelchair' tag in AAT. But we might still question the inclusion of a `wheelchair' tag based on concerns about its \textit{salience} in social and relational contexts. Is the existence of a wheelchair in an image salient enough to be explicitly named? In some cases, the answer may be yes---for instance, in Bennett et al.’s study, the majority of participants \textit{did} describe “disability-related access technologies” in images of themselves \cite{bennett2021s}. But in others, naming the existence of a wheelchair might draw undue and irrelevant attention to the (inferred) disability of a person depicted in an image. Recall the explanation given by Sina Bahram, accessibility consultant for museums, in determining whether to describe wheelchairs in the museum context: “When I think of a picture or a photo of somebody using a wheelchair, I'm not necessarily interested in identifying disability. It would be, however, critical, in certain contexts to know that that person is using a wheelchair and in other contexts it may not be of interest” \cite{TWIMLAI}. Even if we can say with high accuracy and without ontological or epistemological constraints that an object is a wheelchair, salience considerations may caution against their inclusion in AAT in some cases.

These various sources of uncertainty---the technical limitations of computer vision; the ontological instability of identity categories, and epistemological constraints on what can be visually observed about them; and the salience of identity characteristics in relation to the variable social meanings of images, both in their content and in the relational contexts in which they are shared---represent independent considerations that might guide policy choices about how to describe people in alt text. In practice, Facebook (and other companies creating alt text) must be sure neither to conflate these ideas in reasoning about their policies, nor to let resolution on one grounds (e.g., accuracy) be considered normatively sufficient. Rather, each concern must be considered independently when formulating an appropriate alt text policy. 


\section{Navigating uncertainty}

Navigating these different sources of uncertainties is itself an uncertain process. There do not seem to be any obvious or easy answers to the challenges posed by each---and certainly not to all three. We discuss two strategies below: one that has achieved a good deal of traction already (limiting descriptions to only directly observable physical features) and another that is more speculative (relying on facial recognition to name people, rather than describe them)---and the potential limitations and dangers of both approaches.

\subsection{Directly Observable Physical Features}

One approach that might seem to allow Facebook to sidestep some of these concerns is to describe only that which is \textit{visually observable}, rather than imposing identity-related tags---but knowing that some of these descriptions might facilitate inferences about the race, gender, age, disability, etc. of the people so described. For example, rather than attempting to resolve the gender identity of a person that appears in a photo, Facebook could instead describe the clothing, hairstyle, and jewelry that the person is wearing, understanding and accepting that such descriptions may encourage BLV users to draw their own inferences about the person’s gender. This could also happen unintentionally, of course: Facebook may have no explicit goal to facilitate such inferences, but might nonetheless describe these visual features simply in the interest of providing more detailed descriptions. It’s not clear, for example, whether Facebook’s decision to include `beard’ and `jewelry’ \cite{FbookEngBlog} was motivated by any specific concerns with facilitating inference about gender or a desire to provide details beyond gender itself that its annotators deemed especially salient.

Recall that the Cooper Hewitt guidelines suggest such description of visual features in place of gender in some cases: “No assumptions should be made about the gender of a person represented […] When unknown, a person should be described using “they, them” and “person” and their physicality expressed through the description of their features, which inadvertently tend to indicate masculine or feminine characteristics” \cite{CHGuidelines}. To illustrate this principle, the guidelines offer an example of alt text that could be generated for a photograph of people whose genders are unknown, which includes descriptive details that may be read as gender-suggestive: “The person on the left, who is wearing a halter top, leans down to crank a lever as they look over their bare shoulder at the person on the right who pushes their long hair over their shoulder laughing” \cite{CHGuidelines}. Cooper Hewitt here describes certain visual details that annotators themselves may recognize as pertinent to gender expression, while stopping short of ascribing gender to image subjects or intentionally providing clues about gender.

This approach was also favored by nearly all of the interview subjects in Bennett et al.’s study. Of the 25 subjects they interviewed, “24 participants argued AI-generated descriptions would be more respectful if they used appearance rather than identity presumptive language,” with many agreeing that “[a]pproximating skin tone and describing hairstyles may help describers and viewers avoid assuming race; describing clothing, accessories and hairstyles can help describers and viewers avoid gender assumptions, and describing access technologies can help describers and viewers avoid assuming disability” \cite{bennett2021s}. Bennett et al. emphasize that “the preference upon which participants largely converged was that the language of appearance versus that which presumes identity is different” \cite{bennett2021s}.

What might account for this degree of agreement between certain aspects of Facebook’s policy, the museums’ guidelines, and the views expressed by Bennett et al.’s participants? Describing physical characteristics that might serve as the basis for inference about identity categories, but refusing the explicit language of identity categories, seems to strike an interesting balance between competing commitments to autonomy: on the one hand, it still provides people depicted in images with the opportunity to self-define in the identity categories of their own choosing (or to not use any such terms at all), while on the other, it provides BLV users information that might allow them to make their own inferences about people’s identity on the basis of appearance (or to refrain from doing so), just as sighted users could. We might consider it paternalistic for Facebook to deprive BLV users of all visual information that might, for example, insinuate gender---even if that information might serve as the basis for (sometimes incorrect or misguided) inferences, as it might for sighted people. Doing so would deprive BLV users of the opportunity to make their own judgments, leaving it to Facebook to impose ideological constraints on the propriety of inferring personal characteristics from visual data. This reasoning is evident in the argument made by one of Bennett et al.’s research participants: ``If somebody [sighted] sees a photo and has some kind of clues, we should be given comparable information. If you don’t provide that, it does feed into, maybe indirectly, blind people don’t need this information or they don’t make judgements based on this information […W]e’ve gotta find a way to provide it in a way that’s not overly prescriptive’’ \cite{bennett2021s}. Allowing BLV users to draw their own inferences from visual data, rather than Facebook drawing conclusions for them, also permits these inferences to be made individually and privately (i.e., in the heads of BLV users) while Facebook’s might be made in public (i.e., visible to the people being described). This too has normative implications for image subjects: seeing that one has been ascribed a specific gender by Facebook---and knowing that such ascriptions may have been accessed by other Facebook users---can be essentializing and stigmatizing in a way that individual BLV users' inferences might not.

Even when attempting to defer judgment to BLV users, though, Facebook continues to exercise a good deal of influence: the company still determines what visual details to describe in alt text and thus the basis upon which any subsequent inferences might be made by BLV users. The visual cues upon which sighted people rely to draw inferences about identity are numerous, varied, and often subtle---so much so that sighted people might be unable to fully articulate the set of cues upon which they rely when drawing such inferences. Facebook’s model, even with its recently expanded set of concepts, cannot provide descriptions of the full range of visual markers that serve as such cues. Rather, the model is likely to provide details on only fairly obvious and crude cues (e.g., clothing, hairstyle, and jewelry), limiting how discerning any inference might be about the identity categories of the person described in these terms. In this respect, Facebook’s BLV users would not be in the same situation as its sighted users; BLV users’ capacity to infer identity categories---and how such inferences might be drawn---would remain deeply dependent on Facebook’s choices about what descriptors it provides in alt text.\footnote{What, if any, inferences BLV users may draw from these descriptors warrants further study. 
Both \cite{wu2017automatic} and \cite{bennett2021s} repeatedly remark on the skepticism and caution that BLV users exhibit when presented with limited or unreliable information.} In this respect, even though Facebook may be able to defer questions of accuracy, ontology, and epistemology to BLV users themselves by limiting alt text to descriptions of directly observable physical features, the company cannot escape the question of what exactly warrants description and the social meaning of such descriptions. In other words, it cannot avoid questions of salience.

Moreover, in the context of limited information, those descriptors that \textit{are} provided may take on outsized importance. For example, if alt text includes `person' and `red nail polish,’ the nail polish description may be over-read as an indicator of gender, and perhaps reify certain mappings of appearance to gender (even if unreliable). Determinations of what visual descriptors to provide must also consider what information is \textit{not} being provided, and how different visual descriptors might interact with each other. (For example: the gender inference that might be insinuated by `person’ and `long hair’ versus `person,’ `long hair,’ and `guitar’ might differ.) The inferences to be drawn from each descriptor are not made independently.

Finally---and fundamentally---presenting this choice as one between the use of social categories and the description of visual details that might serve as the basis for inference about social categories relies on the idea that there is a clear separation between social categories and directly observable properties. 
Some features are so common in a particular context or among a particular community that they are not even be remarked upon. In a highly homogeneous society, for example, there may be no reason to have concepts concerned with the tone of a person’s skin, the shape of a person’s eyes, or the size of a person’s nose. Their salience as directly observable properties is a result of the fact that they have come to serve as important visual markers for social differentiation. What is “directly observable” encodes particular beliefs about what is an appropriate way to parse the visual world, given the relevance of these features to the social categories that we rely on to make sense of others.


\subsection{Facial Recognition}

Facial recognition might also represent another strategy for navigating these uncertainties. Rather than trying to infer identity categories (`woman’) or trying to describe relevant physical characteristics (`long hair’), those producing alt text could instead try to simply identify the person in question (`Margot Hanley’). Doing so would mean that alt text could report the name of the person, which might allow BLV users to bring prior knowledge to bear about that person’s identity---and perhaps justify providing less or no description of that person’s appearance. In so doing, it might sidestep some of the questions of salience as well as accuracy, ontology, and epistemology, as there would be less or no need to decide which physical features are worthy of description.

In their paper introducing AAT, \cite{wu2017automatic} explain that “while facial recognition turned out to be one of the most requested features from our participants and current technology is viable,” the company was not then prepared to adopt the technique as it had privacy implications for other users. As mentioned, the company has since adopted facial recognition on an opt-in basis, allowing users to elect to be identified automatically in AAT. To our knowledge, Facebook does not use identity to specifically avoid providing other details in its alt text, but it could choose to do so. (Recall that the Cooper Hewitt guidelines warn against this approach, stating that alt text should report identity categories and physical details even when a person has been identified by name.) An alternative approach would be to rely on facial recognition to determine who appears in an image and then match that person’s identity to any information that the person might have shared elsewhere (e.g., in the gender field on Facebook). In that case, the alt text could include the name of the person along with other shared details about their identity categories and appearance.

Despite the potential appeal of facial recognition in order to deal with various sources of uncertainty, there are practical and normative limits to this approach. On social media, naming people, in place of describing their identity or physical appearance, would likely only offer much of value if the people that have been named are known by BLV users.
BLV users may still be interested in the physical appearance of even those with whom they are close (e.g., when a friend changes the color of their hair). Beyond social media and other settings where images largely include family, friends, and associates (e.g., photo management software), the value of this approach will be even more limited, as BLV users will only benefit from facial recognition when the identified person is a public figure. From a normative perspective, this may put other people in an uncomfortable bind, effectively posing opting into facial recognition as a way to forestall the harms that might arise from attempts to describe people in other terms, neither of which they might welcome. Given the serious concerns that BLV users have already expressed about the manual alt text methods that place burdens on others \cite{rzeszotarski2014estimating}, organizations should be careful not to put BLV users in a position where they are made to feel that they are forcing such a decision on their social contacts \cite{Lehrer-Stein_2020}.

\section{Conclusion} 
In this paper, we have explored the tensions that emerge when using computer vision to produce alt text descriptions of people, including identity categories like race, gender, age, disability, etc. We proposed museums as an apt point of comparison, as museums have long navigated these tensions and have developed specific principles and guidelines to aid in their determinations. By comparing the organizations’ policies, we surfaced the normative and practical factors underlying their different approaches. We explained how different forms of uncertainty underlie policy choices about image descriptions, and explored the challenges associated with possible strategies to overcome these.


\section*{Acknowledgements}
This work was supported by the National Science Foundation (CHS-1901151),  the John D. and Catherine T. MacArthur Foundation, and the Brown Institute for Media Innovation. We thank Sina Bahram, Ben Bianchi,  Su Lin Blodgett, A. Feder Cooper, Carter Donohoe, Jacob Ford, Jared Katzman, Kristen Laird, Ruth Starr, Emily Tseng, Hanna Wallach, Meg Young, Freds Madison Avenue, and members of the Artificial Intelligence, Policy, and Practice initiative at Cornell University for valuable discussions.

\bibliographystyle{plain}
\bibliography{bib}

\begin{thebibliography}{10}

\bibitem{facebookai2021}
Facebook AI.
\newblock How facebook is using ai to improve photo descriptions for people who
  are blind or visually impaired, jan 2021.

\bibitem{barlas2019social}
P{\i}nar Barlas, Kyriakos Kyriakou, Styliani Kleanthous, and Jahna Otterbacher.
\newblock Social b(eye) as: Human and machine descriptions of people images.
\newblock In {\em Proceedings of the International AAAI Conference on Web and
  Social Media}, volume~13, pages 583--591, 2019.

\bibitem{barocas2017problem}
Solon Barocas, Kate Crawford, Aaron Shapiro, and Hanna Wallach.
\newblock {The Problem With Bias: Allocative Versus Representational Harms in
  Machine Learning}.
\newblock In {\em Proceedings of SIGCIS}, 2017.

\bibitem{Googleapologises}
BBC.
\newblock Google apologises for photos app's racist blunder, july 2015.

\bibitem{bennett2021s}
Cynthia~L Bennett, Cole Gleason, Morgan~Klaus Scheuerman, Jeffrey~P Bigham,
  Anhong Guo, and Alexandra To.
\newblock “it’s complicated”: Negotiating accessibility and (mis)
  representation in image descriptions of race, gender, and disability.
\newblock 2021.

\bibitem{bennett2019point}
Cynthia~L Bennett and Os~Keyes.
\newblock What is the point of fairness? disability, ai and the complexity of
  justice.
\newblock {\em ACM SIGACCESS Accessibility and Computing}, (125), 2020.

\bibitem{benthall2019racial}
Sebastian Benthall and Bruce~D Haynes.
\newblock Racial categories in machine learning.
\newblock In {\em Proceedings of the conference on fairness, accountability,
  and transparency}, pages 289--298, 2019.

\bibitem{bigham2010vizwiz}
Jeffrey~P Bigham, Chandrika Jayant, Hanjie Ji, Greg Little, Andrew Miller,
  Robert~C Miller, Robin Miller, Aubrey Tatarowicz, Brandyn White, Samual
  White, et~al.
\newblock Vizwiz: nearly real-time answers to visual questions.
\newblock In {\em Proceedings of the 23nd annual ACM symposium on User
  interface software and technology}, pages 333--342, 2010.

\bibitem{birhane2021large}
Abeba Birhane and Vinay~Uday Prabhu.
\newblock Large image datasets: A pyrrhic win for computer vision?
\newblock In {\em Proceedings of the IEEE/CVF Winter Conference on Applications
  of Computer Vision}, pages 1537--1547, 2021.

\bibitem{bowker2000sorting}
Geoffrey~C Bowker and Susan~Leigh Star.
\newblock {\em Sorting things out: Classification and its consequences}.
\newblock MIT press, 2000.

\bibitem{brady2015gauging}
Erin Brady, Meredith~Ringel Morris, and Jeffrey~P Bigham.
\newblock Gauging receptiveness to social microvolunteering.
\newblock In {\em Proceedings of the 33rd Annual ACM Conference on Human
  Factors in Computing Systems}, pages 1055--1064, 2015.

\bibitem{branham2017someone}
Stacy~M Branham, Ali Abdolrahmani, William Easley, Morgan Scheuerman, Erick
  Ronquillo, and Amy Hurst.
\newblock "is someone there? do they have a gun" how visual information about
  others can improve personal safety management for blind individuals.
\newblock In {\em Proceedings of the 19th International ACM SIGACCESS
  Conference on Computers and Accessibility}, pages 260--269, 2017.

\bibitem{buolamwini2018gender}
Joy Buolamwini and Timnit Gebru.
\newblock Gender shades: Intersectional accuracy disparities in commercial
  gender classification.
\newblock In {\em Conference on fairness, accountability and transparency},
  pages 77--91. PMLR, 2018.

\bibitem{CHGuidelines}
Cooper{\ }Hewitt.
\newblock Guidelines for image description, 2021.

\bibitem{crawford2019excavating}
Kate Crawford and Trevor Paglen.
\newblock Excavating ai: the politics of images in machine learning training
  sets.
\newblock {\em Excavating AI}, 2019.

\bibitem{FbookEngBlog}
Manohar~Paluri Darío García~García, Shaomei~Wu.
\newblock Under the hood: Building accessibility tools for the visually
  impaired on facebook, april 2016.

\bibitem{davis2005invisible}
N~Ann Davis.
\newblock Invisible disability.
\newblock {\em Ethics}, 116(1):153--213, 2005.

\bibitem{TechCrunchFacebook}
Megan~Rose Dickey.
\newblock Facebook’s accessibility ambitions, may 2018.

\bibitem{FacebookStats}
Facebook{\ }Inc.
\newblock The{\ }latest{\ }facebook{\ }statistics{\ }(2018), January 2018.

\bibitem{fang2015captions}
Hao Fang, Saurabh Gupta, Forrest Iandola, Rupesh~K Srivastava, Li~Deng, Piotr
  Doll{\'a}r, Jianfeng Gao, Xiaodong He, Margaret Mitchell, John~C Platt,
  et~al.
\newblock From captions to visual concepts and back.
\newblock In {\em Proceedings of the IEEE conference on computer vision and
  pattern recognition}, pages 1473--1482, 2015.

\bibitem{Chancey}
Chancey Fleet.
\newblock Things which garner a ton of glowing reviews from mainstream outlets
  without being of much use to disabled people. for instance, facebook's auto
  image descriptions, much loved by sighted journos but famously useless in the
  blind community, January 2021.

\bibitem{MashableFacebook}
Nicole Gallucci.
\newblock Facebook's new facial recognition efforts help blind users know
  exactly who's in photos, december 2017.

\bibitem{GoogleGender}
Shona Ghosh.
\newblock Google ai will no longer use gender labels like 'woman' or 'man' on
  images of people to avoid bias, feb 2020.

\bibitem{gleason2019s}
Cole Gleason, Patrick Carrington, Cameron Cassidy, Meredith~Ringel Morris,
  Kris~M Kitani, and Jeffrey~P Bigham.
\newblock “it's almost like they're trying to hide it”: How user-provided
  image descriptions have failed to make twitter accessible.
\newblock In {\em The World Wide Web Conference}, pages 549--559, 2019.

\bibitem{GoogleCloudAPI}
Google.
\newblock Vision ai, January 2021.

\bibitem{guinness2018caption}
Darren Guinness, Edward Cutrell, and Meredith~Ringel Morris.
\newblock Caption crawler: Enabling reusable alternative text descriptions
  using reverse image search.
\newblock In {\em Proceedings of the 2018 CHI Conference on Human Factors in
  Computing Systems}, pages 1--11, 2018.

\bibitem{hamidi2018gender}
Foad Hamidi, Morgan~Klaus Scheuerman, and Stacy~M Branham.
\newblock Gender recognition or gender reductionism? the social implications of
  embedded gender recognition systems.
\newblock In {\em Proceedings of the 2018 chi conference on human factors in
  computing systems}, pages 1--13, 2018.

\bibitem{hanley2020ethical}
Margot Hanley, Apoorv Khandelwal, Hadar Averbuch-Elor, Noah Snavely, and Helen
  Nissenbaum.
\newblock An ethical highlighter for people-centric dataset creation.
\newblock {\em arXiv preprint arXiv:2011.13583}, 2020.

\bibitem{hashemi2020criminal}
Mahdi Hashemi and Margeret Hall.
\newblock Criminal tendency detection from facial images and the gender bias
  effect.
\newblock {\em Journal of Big Data}, 7(1):1--16, 2020.

\bibitem{hoffmann2018data}
Anna~Lauren Hoffmann.
\newblock Data violence and how bad engineering choices can damage society.
\newblock {\em Medium. Retrieved June}, 19:2019, 2018.

\bibitem{hoffmann2020terms}
Anna~Lauren Hoffmann.
\newblock Terms of inclusion: Data, discourse, violence.
\newblock {\em New Media \& Society}, pages 1--18, 2020.

\bibitem{huang2016there}
Ting-Hao Huang, Walter Lasecki, Amos Azaria, and Jeffrey Bigham.
\newblock "is there anything else i can help you with?" challenges in deploying
  an on-demand crowd-powered conversational agent.
\newblock In {\em Proceedings of the AAAI Conference on Human Computation and
  Crowdsourcing}, volume~4, 2016.

\bibitem{AmazonRekognition}
Amazon.com Inc.
\newblock Amazon rekognition, January 2021.

\bibitem{jo2020lessons}
Eun~Seo Jo and Timnit Gebru.
\newblock Lessons from archives: Strategies for collecting sociocultural data
  in machine learning.
\newblock In {\em Proceedings of the 2020 Conference on Fairness,
  Accountability, and Transparency}, pages 306--316, 2020.

\bibitem{karpathy2015deep}
Andrej Karpathy and Li~Fei-Fei.
\newblock Deep visual-semantic alignments for generating image descriptions.
\newblock In {\em Proceedings of the IEEE conference on computer vision and
  pattern recognition}, pages 3128--3137, 2015.

\bibitem{kazimzade2020biased}
Gunay Kazimzade and Milagros Miceli.
\newblock Biased priorities, biased outcomes: Three recommendations for
  ethics-oriented data annotation practices.
\newblock In {\em Proceedings of the AAAI/ACM Conference on AI, Ethics, and
  Society}, pages 71--71, 2020.

\bibitem{keyes2018misgendering}
Os~Keyes.
\newblock The misgendering machines: Trans/hci implications of automatic gender
  recognition.
\newblock {\em Proceedings of the ACM on human-computer interaction},
  2(CSCW):1--22, 2018.

\bibitem{kyriakou2019fairness}
Kyriakos Kyriakou, P{\i}nar Barlas, Styliani Kleanthous, and Jahna Otterbacher.
\newblock Fairness in proprietary image tagging algorithms: A cross-platform
  audit on people images.
\newblock In {\em Proceedings of the International AAAI Conference on Web and
  Social Media}, volume~13, pages 313--322, 2019.

\bibitem{Engadget}
Nicole Lee.
\newblock Behind facebook's efforts to make its site accessible to all, july
  2016.

\bibitem{Lehrer-Stein_2020}
Janni Lehrer-Stein.
\newblock What it’s like to use facebook when you’re blind.
\newblock {\em The New York Times}, Jan 2020.

\bibitem{macleod2017understanding}
Haley MacLeod, Cynthia~L Bennett, Meredith~Ringel Morris, and Edward Cutrell.
\newblock Understanding blind people's experiences with computer-generated
  captions of social media images.
\newblock In {\em Proceedings of the 2017 CHI Conference on Human Factors in
  Computing Systems}, pages 5988--5999, 2017.

\bibitem{GoogleAccessibilityTeam}
Dominic Mazzoni.
\newblock Using ai to give people who are blind the “full picture”, dec
  2019.

\bibitem{miceli2020between}
Milagros Miceli, Martin Schuessler, and Tianling Yang.
\newblock Between subjectivity and imposition: Power dynamics in data
  annotation for computer vision.
\newblock {\em Proceedings of the ACM on Human-Computer Interaction},
  4(CSCW2):1--25, 2020.

\bibitem{MicrosoftAzure}
Microsoft{\ }Azure{\ }Computer{\ }Vision, January 2021.

\bibitem{Collection}
Museum{\ }of{\ }Contemporary{\ }Art{\ }Chicago.
\newblock Collection, 2021.

\bibitem{MCAGuidelines}
Museum{\ }of{\ }Contemporary{\ }Art{\ }Chicago.
\newblock Mca{\ }guidelines{\ }for{\ }describing, 2021.

\bibitem{VergeFacebook}
Casey Newton.
\newblock Facebook begins using artificial intelligence to describe photos to
  blind users, april 2016.

\bibitem{otterbacher2018social}
Jahna Otterbacher.
\newblock Social cues, social biases: stereotypes in annotations on people
  images.
\newblock In {\em Proceedings of the AAAI Conference on Human Computation and
  Crowdsourcing}, volume~6, 2018.

\bibitem{otterbacher2019we}
Jahna Otterbacher, P{\i}nar Barlas, Styliani Kleanthous, and Kyriakos Kyriakou.
\newblock How do we talk about other people? group (un) fairness in natural
  language image descriptions.
\newblock In {\em Proceedings of the AAAI Conference on Human Computation and
  Crowdsourcing}, volume~7, pages 106--114, 2019.

\bibitem{LiorTheFix}
Alica Peszkowska.
\newblock Moving communities online (without losing their substance), march
  2020.

\bibitem{Wapo}
Andrea Peterson.
\newblock How facebook is helping the blind ‘see’ pictures their friends
  share online, april 2016.

\bibitem{petrie2005describing}
Helen Petrie, Chandra Harrison, and Sundeep Dev.
\newblock Describing images on the web: a survey of current practice and
  prospects for the future.
\newblock {\em Proceedings of Human Computer Interaction International (HCII)},
  71:2, 2005.

\bibitem{MicrosoftAutoDescriptions}
John Roach.
\newblock What’s that? microsoft’s latest breakthrough, now in azure ai,
  describes images as well as people do, 2020.

\bibitem{rzeszotarski2014estimating}
Jeffrey~M Rzeszotarski and Meredith~Ringel Morris.
\newblock Estimating the social costs of friendsourcing.
\newblock In {\em Proceedings of the SIGCHI Conference on Human Factors in
  Computing Systems}, pages 2735--2744, 2014.

\bibitem{salisbury2017toward}
Elliot Salisbury, Ece Kamar, and Meredith Morris.
\newblock Toward scalable social alt text: Conversational crowdsourcing as a
  tool for refining vision-to-language technology for the blind.
\newblock In {\em Proceedings of the AAAI Conference on Human Computation and
  Crowdsourcing}, volume~5, 2017.

\bibitem{scheuerman2019computers}
Morgan~Klaus Scheuerman, Jacob~M Paul, and Jed~R Brubaker.
\newblock How computers see gender: An evaluation of gender classification in
  commercial facial analysis services.
\newblock {\em Proceedings of the ACM on Human-Computer Interaction},
  3(CSCW):1--33, 2019.

\bibitem{scheuerman2020we}
Morgan~Klaus Scheuerman, Kandrea Wade, Caitlin Lustig, and Jed~R Brubaker.
\newblock How we've taught algorithms to see identity: Constructing race and
  gender in image databases for facial analysis.
\newblock {\em Proceedings of the ACM on Human-Computer Interaction},
  4(CSCW1):1--35, 2020.

\bibitem{annatweet}
Anna Selle.
\newblock creepy that facebook is generating this, but i think it’s just alt
  text for screen readers. so probably just an accessibility thing, but also
  unsettling if that text isn’t being generated directly by users., July
  2019.

\bibitem{stangl2020person}
Abigale Stangl, Meredith~Ringel Morris, and Danna Gurari.
\newblock "person, shoes, tree. is the person naked?" what people with vision
  impairments want in image descriptions.
\newblock In {\em Proceedings of the 2020 CHI Conference on Human Factors in
  Computing Systems}, pages 1--13, 2020.

\bibitem{tran2016rich}
Kenneth Tran, Xiaodong He, Lei Zhang, Jian Sun, Cornelia Carapcea, Chris
  Thrasher, Chris Buehler, and Chris Sienkiewicz.
\newblock Rich image captioning in the wild.
\newblock In {\em Proceedings of the IEEE conference on computer vision and
  pattern recognition workshops}, pages 49--56, 2016.

\bibitem{TWIMLAI}
TWIMLFest.
\newblock Accessibility and computer vision, oct 2020.

\bibitem{van2016stereotyping}
Emiel Van~Miltenburg.
\newblock Stereotyping and bias in the flickr30k dataset.
\newblock {\em arXiv preprint arXiv:1605.06083}, 2016.

\bibitem{van2018talking}
Emiel van Miltenburg, Desmond Elliott, and Piek Vossen.
\newblock Talking about other people: an endless range of possibilities.
\newblock In {\em Proceedings of the 11th International Conference on Natural
  Language Generation}, pages 415--420, 2018.

\bibitem{VergeOutage}
James Vincent.
\newblock Facebook’s image outage reveals how the company’s ai tags your
  photos, july 2019.

\bibitem{MykalTweet}
Mykal Vincent.
\newblock Facebook what up with all these broken images and alt text revealing
  your creepy crawly algorithm terms??? @facebook, July 2019.

\bibitem{Artsy}
Claire Voon.
\newblock This open source software could make museum websites more accessible,
  2019.

\bibitem{wang2018deep}
Yilun Wang and Michal Kosinski.
\newblock Deep neural networks are more accurate than humans at detecting
  sexual orientation from facial images.
\newblock {\em Journal of Personality and Social Psychology}, 114(2):246, 2018.

\bibitem{whittaker2019disability}
Meredith Whittaker, Meryl Alper, Cynthia~L Bennett, Sara Hendren, Liz Kaziunas,
  Mara Mills, Meredith~Ringel Morris, Joy Rankin, Emily Rogers, Marcel Salas,
  et~al.
\newblock Disability, bias, and ai.
\newblock {\em AI Now Institute, November}, 2019.

\bibitem{wu2017automatic}
Shaomei Wu, Jeffrey Wieland, Omid Farivar, and Julie Schiller.
\newblock Automatic alt-text: Computer-generated image descriptions for blind
  users on a social network service.
\newblock In {\em Proceedings of the 2017 ACM Conference on Computer Supported
  Cooperative Work and Social Computing}, pages 1180--1192, 2017.

\bibitem{yang2020towards}
Kaiyu Yang, Klint Qinami, Li~Fei-Fei, Jia Deng, and Olga Russakovsky.
\newblock Towards fairer datasets: Filtering and balancing the distribution of
  the people subtree in the imagenet hierarchy.
\newblock In {\em Proceedings of the 2020 Conference on Fairness,
  Accountability, and Transparency}, pages 547--558, 2020.

\bibitem{zhao2017understanding}
Yuhang Zhao, Michele Hu, Shafeka Hashash, and Shiri Azenkot.
\newblock Understanding low vision people's visual perception on commercial
  augmented reality glasses.
\newblock In {\em Proceedings of the 2017 CHI Conference on Human Factors in
  Computing Systems}, pages 4170--4181, 2017.

\bibitem{zhao2017effect}
Yuhang Zhao, Shaomei Wu, Lindsay Reynolds, and Shiri Azenkot.
\newblock The effect of computer-generated descriptions on photo-sharing
  experiences of people with visual impairments.
\newblock {\em Proceedings of the ACM on Human-Computer Interaction},
  1(CSCW):1--22, 2017.

\end{thebibliography}

\end{document}